\documentclass{iopart}
\usepackage{graphicx}

\eqnobysec

\def\OMIT#1 {{}}
\def\MEMO#1 {{}}
\newcommand{\eqr}[1]{(\ref{#1})}
\newcommand{\half}{{\textstyle{1\over2}}}
\newcommand{\bbeta}{\beta}  %%% edit this to LOWER \beta in subscripts

\newcommand {\vdimer}    {{\vrule height0.2cm width0.05cm depth0pt}}
\newcommand {\hdimer}    {{\hrule height0.05cm width0.2cm depth0pt}}
\newcommand {\vdimers}   {\hbox{\vdimer \hskip 0.1cm \vdimer}}
\newcommand {\hdimers}   {\hbox{\vbox{\hdimer \vskip 0.1cm \hdimer}}}
\newcommand {\updown}    {\uparrow\downarrow}
\newcommand {\downup}    {\downarrow\uparrow}
\newcommand {\HH} {{\cal H}}
\newcommand {\Hclass} {{\overline{\HH}}}
\newcommand {\Ns}{{\cal {N}_{\rm s}}}
\newcommand {\Nflip}{{\cal {N}_{\rm flip}}}
\newcommand {\ZZ} {{\cal Z}}

\newcommand {\thop}{t}
\newcommand {\Vdim}{V}
\newcommand {\ttime}{\tau}
\newcommand {\tA}{{\tilde A}}
\newcommand {\tB}{{\tilde B}}
\newcommand{\Pzero}{P^{(0)}}
\newcommand{\philambda}{\phi^{(\lambda)}}
\newcommand{\Cquant}{C}
\newcommand{\Cclass}{{C^{\rm class}}}
\newcommand{\HHFM}{{\HH_{\rm FM}}} 
\newcommand{\tW}{{\tilde W}}

\newcommand {\rr} {{\bf r}}
\newcommand {\qq} {{\bf q}}
\newcommand {\la} {\langle}
\newcommand {\ra} {\rangle}

\begin{document}

\title
[Classical to quantum dynamics at Rokhsar-Kivelson points]
{From classical to quantum dynamics at Rokhsar-Kivelson points}

\author{C.~L.~Henley}

\address{Department of Physics, 
Cornell University, Ithaca, New York 14853-2501, USA}

\begin{abstract}
For any classical statistical-mechanics model with a discrete state space, 
and endowed with a dynamics satisfying detailed balance, it is possible
to generalize the Rokhsar-Kivelson point for the quantum dimer model.
That is, a quantum Hamiltonian can be constructed (on the same state
space), such that the ground state wavefunction coincides with the
classical equilibrium distribution. Furthermore the excited eigenstates
correspond to classical relaxation modes, which (in cases with a
symmetry or conserved quantity) permits extraction of the dispersion
law of long-wavelength excitations.  The mapping is natural mainly
when the states have equal weight, as is typical of a highly frustrated
model. Quantum and classical correlation functions
are related by analytic continuation to the imaginary time axis. 
\end{abstract}
%%% \pacs{71.10.Fd, 71.10.Pm, 05.30.Jp, 74.20.Mn}

%% \maketitle

\section{introduction}
\label{sec:intro}

\MEMO{Edit defn of $\bbeta$ to lower it (used in subscripts only)}.

The quantum dimer model~\cite{qdm} of Rokhsar and Kivelson (RK)
was inspired by the resonating-valence bond state
of a quantum antiferromagnet,  which had just been 
proposed as a starting point for the explanation of
high-temperature superconductivity.  That was not to
be, but models of this type are of interest among
the ``highly frustrated systems''~\cite{Moe01b} 
characterized by massive degeneracy (or near-degeneracy). 
Generalizations of the RK construction were used to construct
the first concrete lattice models in dimension $d>1$
that manifestly exhibit fractionalized 
excitations~\cite{Moe01a,Moe01b,Na01,Ba01,Mo02}
\MEMO{Check: is \cite{Mo02} in this class?}

The quantum dimer model had one nontrivial parameter $\Vdim/\thop$
(see Eq.~\eqr{eq:HQDM}, below);
when it takes a special value (``RK point''), 
RK showed the exact ground state wavefunction is an equal-weighted
superposition of all dimer coverings. This had the same
(critical) static correlations as the classical dimer 
ensemble, an exactly solved model~\cite{fisher,fisher-st}. 
Thus the RK point is a rather special kind of quantum critical point.
\MEMO{unusual in that it is understood better than the phases it separates.}

Later the author noticed that, at the RK point, 
{\it excited} eigenstates correspond exactly to relaxation modes
of the master equation for the natural Monte Carlo (MC) dynamics
of the classical dimer ensemble~\cite{dydim,statphys}.
Furthermore, the dimer ensemble (on bipartite lattices) has
a natural coarse-graining via the height representation~\cite{Bl02}, 
whereby it maps to a 2+1 dimensional interface model in its rough phase.
Since the classical dynamics is easy to grasp, this mapping
delivered the dispersion law for the quantum dimer model's
elementary excitations at the RK point.
(This dispersion was already understood variationally~\cite{qdm}).

The purpose of this paper is to generalize the RK construction to
any (degenerate) classical ensemble, showing the correspondence
for several simple classical models. 
I begin (Sec.~\ref{sec:QDM})
by reviewing basic notions of the quantum dimer model\cite{qdm,dydim}
including the way that the classical and quantum models are connected, 
and the way we can comprehend the quantum dispersion if we know
the classical dispersion.  
%%% In Sec.~\ref{sec:stategraph}. 
%%% the ``state graph'' picture is introduced
%%% as an explicit way to think of these many-body problems as a sort of diffusion
%%% (or localization) of a single particle
%%% on a graph in some abstract or high-dimensional space.}
Any discrete classical model 
(e.g. ~\cite{Pa02}) 
with a dynamics satisfying detailed balance can be 
``Rokhsar-Kivelsonized'' to produce a quantum model
with the same mapping of the eigenfunctions to 
classical dynamics (Sec.~\ref{sec:gen}).

Some examples are 
(i) Chakravarty's quantum 6-vertex model~\cite{q6v}.
(ii) A classical Ising chain with spin-exchange
(Kawasaki) dynamics, for which 
the RK model is the spin-1/2 Heisenberg ferromagnet.
(iii) A spin-1/2 Ising model on the pyrochlore lattice~\cite{hermele}.
The examples are 
built on a large basis set of essentially degenerate states.
In that sense they are ``highly frustrated'' models~\cite{ramirez}, 
whether or not the massive degeneracy arises from competing interactions.

Finally (Sec.~\ref{sec:dyn})
it is verified that the classical and quantum 
correlation functions are related exactly by a rotation
of real time into imaginary time (and thus might be
extracted more easily from simulations).

\section{Classical and quantum dimer model}
\label{sec:QDM}

\MEMO{
The QDM Hamiltonian has a ``kinetic'' amplitude $-\thop$ for each possible 
``flip'' (between the two ways to arrange dimers on opposite sides 
of a square), and a ``potential'' term with a cost $\Vdim$ for each 
flippable plaquette.}

The  Hilbert space of the quantum dimer model consists of all complete 
dimer covering configurations.
Its Hamiltonian is customarily written
  \begin{equation}
     \HH = - \thop \sum \Bigl( | \vdimers \rangle \langle \hdimers| + 
           {\rm h.c.} \Bigr) + \Vdim \sum \Bigl( |\vdimers \rangle \langle \vdimers|
      + |\hdimers \rangle \langle \hdimers| \Bigr). 
  \label{eq:HQDM}
  \end{equation}
for a square lattice. 
In the $\thop$ term, by an abuse of notation, ``$|\vdimers \rangle \langle \hdimers|$''
actually runs over $|\beta\rangle \langle \alpha|$ 
for every possible pair of configurations 
$(\beta, \alpha)$, such that $\beta$ differs from $\alpha$ only by the 
replacement of a vertical pair by a horizontal pair of dimers on one plaquette. 
This is the elementary ``flip move'' for this model -- the smallest possible
change that turns one valid configuration into a different one (since the same 
four vertices are covered in either state).   
(For a dimer covering on a general bipartite lattice, flippable
plaquettes are those around which every second edge has a dimer, 
and the flip move exchanges covered and uncovered edges.)
The $\Vdim$ term is diagonal in this
Hilbert space, and can be rewritten $\Vdim {\Nflip}$, where 
$\Nflip(\alpha)$ (dependent on
the configuration) is the number of flippable plaquettes in configuration $\alpha$.

RK noted that when $\Vdim=\thop$ (the ``RK point''), the ground state wavefunction is
   \begin{equation}
        |\Psi_0\rangle = \frac{1}{\sqrt{\Ns}}  \sum_\alpha |\alpha\rangle, 
   \label{eq:Psiground}
   \end{equation}
where the sum is over all valid dimer configurations.
Thus, the probability weight is $\rm{Prob}(\alpha) = \Ns^{-1}$,  the same for
each of the $\Ns$ states, just as in the classical ensemble.

Any discrete classical model's dynamics is described by the master equation
   \begin{equation}
        \dot p_\alpha(\ttime) = 
           \sum _ {\beta [\beta\neq \alpha]} 
           \Bigl( W_{\alpha \beta} p_\bbeta(\ttime)
            - W_{\beta \alpha} p_\alpha(\ttime) \Bigr). 
   \label{eq:master}
  \end{equation}
Here $p_\alpha(\ttime)$ is the instantaneous probability to be
in configuration $\alpha$ at time $\ttime$, and $W_{\alpha\bbeta}$ is the transition
rate to state $\alpha$, given the system is in state $\beta$.
%%  of course the two terms in \eqr{eq:master} are respectively 
%% the scatterings into and out of state $\alpha$.  

If we define 
   \begin{equation}
          W_{\alpha\alpha} \equiv - \Gamma_\alpha \equiv 
           - \sum _ {\beta [\beta\neq \alpha]} 
           W_{\beta\alpha},
   \end{equation}
(total transition rate out of $\alpha$), 
then \eqr{eq:master} can be rewritten as a matrix equation, 
   \begin{equation}
          \dot {\bf p}(\ttime) = {\bf W} {\bf p}(\ttime).
   \end{equation}
Then the time evolution can obviously be decomposed into eigenmodes
of the $\bf W$ matrix, labeled by eigenvalue $-\lambda$ where $\lambda \ge 0$:
   \begin{equation}
          p_\alpha(\ttime) = \sum_\lambda c_\lambda  \rme^{-\lambda \ttime} 
          \phi^{(\lambda)}_\alpha
   \label{eq:timeevol}
   \end{equation}
where $ \phi^{(\lambda)}_\alpha$ is the (normalized) eigenvector.

We actually are concerned only with the states that are connected by flips
to a given initial state;  within this component we can invoke a 
variant of the Perron-Frobenius theorem  to assert that there is
a unique steady state distribution, $P^{(0)}_\alpha$, characterized in
matrix notation by ${\bf W} {\bf P}^{(0)} = 0$ (this distribution is
just $\phi^{(0)}$ with the sum of the components normalized to unity, 
not the sum of squares.)

Furthermore, say the classical model has a Hamiltonian 
$\Hclass (\alpha)$ (we take a dimensionless $\Hclass$ which 
has already been divided by the temperature). 
Then the steady state should be the Boltzmann distribution, 
   \begin{equation}
          P^{(0)}_\alpha = \rme^ {-\Hclass(\alpha)}/ \ZZ, 
   \label{eq:boltzmann}
   \end{equation}
where $\ZZ \equiv \sum_\alpha \rme^{-\Hclass(\alpha)}$ is the partition function.
Observe that $\HH$ and $\Hclass$ are {\it not} related, at least
not in the usual sense of taking the classical limit of a quantum
dynamics. 

Of course, the 
dynamics should satisfy detailed balance, 
   \begin{equation}
          W_{\beta\alpha} P^{(0)}_\alpha = W_{\alpha\bbeta} P^{(0)}_\bbeta.
   \label{eq:detbal}
   \end{equation}

For the quantum dimer model, and also the generalized Rokhsar-Kivelsonized models, 
we will specialize to 
   \begin{equation}
          \Hclass =0, 
   \label{eq:zeroH}
   \end{equation}
i.e. the allowed configurations
are all degenerate.  In this case, $P^{(0)}(\alpha)$ is the same
for every $\alpha$ and so \eqr{eq:detbal} reduces to 
saying $W_{\beta\alpha} = W_{\alpha\bbeta}$, i.e. the rate matrix
is symmetric. 
Next observe that if the classical model's flip rate is always $w$ 
(whenever a flip is possible), i.e.  $W_{\alpha \beta} = w$ or $0$, 
then the matrix elements from \eqr{eq:HQDM} are
$\HH_{\alpha\bbeta} = -\frac{\thop}{w} W_{\alpha\bbeta} + 
            \frac{\Vdim}{w} \Gamma_\alpha \delta_{\alpha\bbeta}$.
Thus, at the RK point, 
the quantum Hamiltonian matrix is proportional
to the classical rate matrix:
   \begin{equation}
       \HH  \equiv - \frac{\thop}{w} {\bf W}.
   \label{eq:classquantum}
   \end{equation}
\OMIT{Evidently, since $\HH$ is Hermitean, this requires the Hamiltonian
be degenerate as in Eq.~\eqr{eq:zeroH}, so that ${\bf W}$ is symmetric.}
%%%%%%%%%%%%%%%%%%%%%%
It follows, 
%%%% from \eqr{eq:classquantum},  
of course, that all the 
eigenvectors of the quantum matrix are the same as
those of the classical matrix, and the eigenenergies are given by 
   \begin{equation}
        E_\lambda= \frac{\thop}{w} \lambda .
   \label{eq:E-lambda}
   \end{equation}
This is the key result of the present paper. 
The quantum ground state eigenfunction \eqr{eq:Psiground} 
is just the special case which has $\lambda=0$:
its identity with the classical stationary state, noted originally 
by RK~\cite{qdm},  follows since ${\bf P}^{(0)}$ is a null vector of
${\bf W}$.

\section{Generalizations and examples}
\label{sec:gen}

Now consider {\it any} classical model with discrete configurations 
\OMIT{(e.g.  an Ising model, a tiling,  or a lattice gas)}
with null Hamiltonian \eqr{eq:zeroH}. 
Define a set of allowed ``flips'' (connecting two configurations)
and endow the model with a Monte Carlo dynamics 
(in continuous time) such that every possible flip has rate $w$.
Furthermore, on the Hilbert space $\{ |\alpha\ra \}$, 
define a quantum-mechanical 
Hamiltonian that includes those same flips (with amplitude $\thop$), 
as well as a term $\Vdim \Nflip $ that penalizes a configuration
once for every possible flip move.
At the RK  point $\thop=\Vdim$, the Hamiltonian and 
master-equation matrices are proportional, so as before
the eigenvectors are the same and \eqr{eq:E-lambda} holds.

In many cases, one can construct a coarse-grained field 
from these configurations and  infer the relaxational dynamics 
of the classical model at long wavelengths.
Thus by the mapping, one also understands the low-energy 
excited states of the quantum model, 
a nontrivial problem if it were approached directly.

\subsection {Discrete models with ``height'' representations}

The quantum dimer model belongs to a class of
models that are coarse-grained via a microscopic mapping
of each microstate to an interface $\{ h(\rr) \}$
in an abstract $2+1$ dimensional space~\cite{dydim,Bl02,zeng}.
The interface is in its roughened phase, so the classical model
is described by an effective free energy density $\propto |\nabla h(\rr)|^2$.
The normal modes of the classical master equation 
are simply capillary modes of this interface
with eigenvalues $\lambda ({\bf q})  \propto |\qq|^2$. 
Hence~\cite{dydim,qdspin} the quantum excitations are bosons with 
dispersion 
   \begin{equation}
       \hbar \omega (\qq) \propto |\qq|^2 .
   \label{eq:q2disp}
   \end{equation}

Height models have conserved quantities, often called ``winding numbers'' 
but most transparently understood as the components $(\nabla_x h, \nabla_y h)$. 
Local flip moves cannot change the global ``interface'' slope $\nabla h$.
Hence the configuration space is partitioned into subspaces,  each of which has 
a steady state under the classical dynamics and a corresponding 
RK quantum ground state with zero energy.  
So, just as a Goldstone mode follows from a symmetry, 
the dispersion \eqr{eq:q2disp} 
is related to the degeneracy under changes of $\nabla h$.

The earliest quantum height model ({\it not} then recognized as such)
was the Anderson-Fazekas ~\cite{fazekas} approach to
the s=1/2 triangular lattice antiferromagnet  from the Ising limit.
The basis is the Ising ground states, which essentially map to
the dimer coverings of a honeycomb lattice.  (The triangular
Ising antiferromagnet in a transverse field maps directly to
the honeycomb quantum dimer model~\cite{Moe01b}.
A more interesting spin model that maps to the square-lattice
quantum dimer model is the 
nearly Ising $s=1/2$ antiferromagnet on the
``checkerboard'' lattice~\cite{Moe01b,qdspin}.)

S. Chakravarty has introduced a ``quantum six-vertex'' (Q6V) 
model~\cite{q6v}, a generalization of a speculative state in which 
spontaneous orbital currents develop along the lattice edges,
as was proposed for the pseudogap phase of cuprates.
The currents define arrow variables, which satisfy an ice rule at
vertices. 
\OMIT{the degrees of freedom are arrows on the edges of
a square lattice; configurations are allowed if every vertex has two
inward and two outward arrows.}  
The minimal flip move is to reverse 
all four arrows around a plaquette, 
provided they all point in the same clock sense.~\footnote{
%%%%%%%%%%%%%%%
This is equivalent to the transverse field with 
projector in (16) and (17) of Ref.~\cite{q6v}.}
%%%%%%%%%%
The 6-vertex configurations have a height represensation~\cite{vB77}, 
and the whole $T=0$ behavior is strictly parallel to the QDM of RK~\cite{qdm}
(assuming that analogous terms are included in the Hamiltonian!)
In particular, when $\Vdim=0$, a ``flat'' phase with gapped excitations
occurs;
in it, the arrows has long-range alternating order of the so-called
``d-density wave'' type~\cite{q6v}, though with fluctuations.
At the RK point $\Vdim=t$, however, a critical phase is predicted
with dispersion \eqr{eq:q2disp}.
Quantum height models at the RK point will be further discussed
%% onlinecite
in \cite{qdspin}. 

\OMIT {it is the BCSOS (``body-centered solid-on-solid'') surface model
The flip move corresponds to adding or removing one atom from the
(100) surface of a bcc crystal, which is possible if and only if the four
surrounding atoms all have the same height.}

\subsection {Spin exchange}
\label{sec:spinmodel}

Say that our classical model is a set of Ising spins on a lattice
of $N$ sites in any dimension, with a zero Hamiltonian.
Adopt the spin-conserving ``Kawasaki'' dynamics, i.e. the flip
move is to exchange any nearest-neighbor pair.  
%%%%%%%%%%%%%

It might appear that this example is so trivial that it is
not worth the observation that it may be considered a 
Rokhsar-Kivelson model.  Some of the reasons why it may be
of interest are (i) it adds weight to the conjecture that the
RK point is typically a critical point and that the dispersion
there is generically $q^2$ if there is a conserved quantity;
(ii) a one-dimensional chain of this sort is the simplest
example of the extension of the RK idea to the case
that the classical states are unequally weighted
(Sec.~\ref{sec:weighted}, below);
(iii) the kagom\'e quantum dimer model of Misguich 
{\em et al}~\cite{misguich} maps to this model (see below).

The corresponding spin-$1/2$ quantum Hamiltonian is
   \begin{equation}
       \HHFM = -\thop \sum \Bigl( |\updown\rangle \langle \downup| + {h.c.} \Bigr)
   + \Vdim \sum \Bigl(|\updown \rangle \langle \updown| + 
      |\downup \rangle \langle \downup| \Bigr).
   \label{eq:spinham}
   \end{equation}
(with same abuse of notation as in \eqr{eq:HQDM}).
Converting to the notation of spin operators, each term becomes
   \begin{equation}
-t(S_i^+ S_j^- + S_i^- S_j^+) + \Vdim \Bigl[ 
(\half + S_i^z)(\half - S_j^z) + (\half - S_i^z)(\half + S_j^z) \Bigr] , 
   \end{equation}
%%%%%%
\newcommand{\Jperp}{{J_\perp}}
\newcommand{\Jz}{{J_z}}
where $i,j$ are nearest neighbors; thus
   \begin{equation}
    \HHFM =   \sum _{\langle i j \rangle} - \Jperp (S_i^x S_j^x + S_i^y S_j^y)
           - \Jz (S_i^z S_j^z - \frac{1}{4}), 
   \label{eq:HFM}
   \end{equation}
where $\Jz\equiv 2V$ and $\Jperp\equiv 2t$.
This is the ferromagnet with ``XXZ'' exchange anisotropy 
and the RK point here is the isotropic Heisenberg chain.

Let's adapt \eqr{eq:Psiground} to this case.
The sum over states runs over every possible sequence of up or down spins;
thus the $\Ns=2^N$
terms can be grouped as a direct product of single-site terms, namely
   \begin{equation}
       |\Psi_0\rangle =  \left( \frac{|\uparrow\rangle + |\downarrow\rangle}{2}\right)^N
     = |\rightarrow \, \rightarrow \, \rightarrow \ldots 
        \rightarrow \,\rightarrow \,\rangle, 
   \label{eq:coherent}
   \end{equation}
the ferromagnetic state with moments aligned in the $+x$ direction.

Since the model is isotropic, 
we know there are degenerate ferromagnetic states with moments in other 
directions.~\footnote{
%%%%%%%%%%%%%%%%%
For an alternative viewpoint on $|\Psi_0\ra$, note that
\eqr{eq:detbal} is perfectly compatible with $\Hclass = -h \sum _i S_i^z$
in place of \eqr{eq:zeroH}): then $\Pzero(\alpha)$ directly
maps to \eqr{eq:PsiFMgeneral}; in general, $\Hclass$ could
be any conserved quantity.}
\OMIT{The classical ensemble with probability weight $\rme^{-hM}$
maps directly into \eqr{eq:PsiFMgeneral}) with $\tan \half\theta = \rme^h$.
The most general ground-state wavefunction is $\Psi(\alpha) = \rme^{(-h+i\phi)M/2}$, 
where $\phi$ is the rotation around the $z$ axis (for a real wavefunction, 
the spin can only be in the $xy$ plane).}
%%%%%%%%%%%%%%%%%
This model conserves spin, so the quantum ground state is
degenerate, as explained at the end of Sec.~\ref{sec:QDM}. 
Indeed, the sum in a wavefunction \eqr{eq:Psiground}
should only run over the mutually accessible configurations 
with $(1+\cos\theta)N/2$ up-spins  and $(1-\cos\theta)N/2$ down-spins. 
In the thermodynamic limit this is essentially the direct product 
generalizing \eqr{eq:coherent}, 
   \begin{equation}
      \Bigl(\cos \half\theta |\uparrow\rangle + 
        \sin \half\theta |\downarrow\rangle \Bigr)^N, 
   \label{eq:PsiFMgeneral}
   \end{equation}
a coherent state in which all spins are rotated 
from the $z$ axis by an angle $\theta$.

Now, the classical relaxation dynamics is simple diffusion.
The diffusion constant is easily obtained if we re-imagine the
dynamics as exchanging {\it all} neighbor pairs of spins
at a rate $w$.  (If both spins point the same way, this has no
effect.)  For example, on a chain a marked spin executes a simple random walk
with a total hopping rate of $w$ to the right 
and $w$ to the left.  The long-wavelength behavior is
diffusion, 
      ${d}\sigma(x,\ttime)/d \ttime  =
        D {\partial^2} \sigma(x,\ttime) {\partial x^2}, $
%%     \label{eq:spindiffusion}
where $\sigma(x,\ttime)$ is the spin density, 
and $D\equiv w$.  The eigenvalues of 
this diffusion equation 
%%% \eqr{eq:spindiffusion} 
are $\lambda(q) \cong Dq ^2 \equiv w q^2$ 
for small wavevectors $q$. Hence, via  \eqr{eq:classquantum},  
the quantum model's excitations  have dispersion
       $\hbar \omega(q) \cong  t q^2  \equiv \half \Jperp q^2$.
%%%     \label{eq:spinwaves}
But this is just the familiar formula for ferromagnetic spin waves!

The same classical model, if endowed with 
a {\it single} spin flip dynamics, maps under
Rokhsar-Kivelsonization to 
noninteracting spins in a transverse field --
a trival, gapped quantum model.
Now, one of the most interesting RK models is that
%%% onlinecite
of Ref.~\cite{misguich}. 
Its Hilbert space consists of all dimer coverings of the 
kagom\'e lattice, which (it has been shown)  correspond
one-to-one to the possible $s_z$ spin configurations on the 
triangular Bravais lattice (modulo a global spin reversal). 
%%% onlinecite
Every ``hop'' in the Hamiltonian of \cite{misguich}
rearranges dimers around one hexagon, and this 
simply corresponds to flipping exactly one Ising spins:
this model is precisely that transverse-field model.
(Since every state is flippable, the $V$ term is
trivial in this case).

The above insight, that \eqr{eq:spinham} is the RK map
of the classical ferromagnet, suggests a modification of
the Kagom\'e dimer model, by adopting a hop move
that {\it exchanges} two ``Ising'' spins (corresponding
to a dimer rearrangement around two hexagons). 
But -- since the conservation of $S_z$ has no particular 
meaning in the dimer geometry -- we could make a 
similarity transform $\HH_{FM} \to U \HH _{FM} U$, 
where $U$ is $S_{j}^y$ for a particular $y$, or a
product of such factors for any set of sites. 
Thus one has a family of $2^N$ distinct quantum dimer
models, each with a hop move rearranging two hexagons, 
and each with {\it gapless} excitations  labeled by a
$q^2$ dispersion; it might be interesting to investigate
these models. 

A similar model to \eqr{eq:spinham} that may be Rokhsar-Kivelsonized is 
a classical,  noninteracting lattice gas. 
In the same fashion as the above spin model, it maps to
a quantum model of hard-core bosons with a nearest-neighbor attraction $\Vdim$.
In fact it {\it is} that spin model, 
using a well-known correspondence in which 
up (resp. down) spins are transcribed to occupied (vacant) sites.

\subsection{Pyrochlore model}

Ref.~\cite{hermele} have developed a quantum spin-1/2 model on the
pyrochlore lattice, which is the Rokhsar-Kivelsonization of the
ground state ensemble of the pyrochlore Ising antiferromagnet.
It is well known that those configurations map to those of the
diamond-lattice ice model, with arrows along lattice edges.
The appropriate order parameter for long wavelengths is the 
polarization, the coarse-graining of the ice-model arrow field, 
which is analogous to $\nabla h$ in the height models and is
conserved in the dynamics.  Thus the classical dynamics is
described by a diffusion equation and the dispersion is
\eqr{eq:q2disp} at the RK point.  This contrasts with 
the $\omega \sim |\qq|$ dispersion elsewhere in the phase diagram, 
which Ref.~\cite{hermele} have called ``light.''
It indicates that, even in this model, the RK point is a quantum
critical point.
The $q^2$ dispersion is generic at RK points, if there is a conserved
quantity in the classical model, even in the absence of a height representation.

\MEMO{Senthil's model:
Check it out.  Mention that the models with fractionalized
excitations are NON height models.}

\subsection{Classical dynamics with nontrivial Hamiltonian}
\label{sec:weighted}

The RK mapping is possible even when the classical
ensemble has unequal weights.  The matrix $\tilde{\bf W}$
having elements
  \begin{equation}
      \tW_{\alpha\bbeta} \equiv 
       {\Pzero_\alpha}^{-1/2} W_{\alpha\bbeta} {\Pzero_\bbeta}^{1/2}
       \equiv 
      W_{\alpha\bbeta} \rme^{\half [\Hclass(\alpha)-\Hclass(\beta)]}
  \label{eq:Wsymm}
  \end{equation}
is symmetric, on account of detailed
balance \eqr{eq:detbal}; furthermore, since $\tilde{\bf W}$ is a similarity
transform of $\bf W$, they share the same eigenvalues.  The eigenvectors
\newcommand{\tphilambda}{{\tilde \phi}^{(\lambda)}}
are related by  
%%% $\tphilambda_\alpha \equiv {\Pzero_\alpha}^{-1/2} \philambda_\alpha$, 
$\tphilambda_\alpha \equiv 
\philambda_\alpha / \sqrt{\Pzero_\alpha}$, 
where $\philambda$ refers to a {\it right} eigenvector of $\bf W$. 
The corresponding quantum Hamiltonian matrix must be proportional 
to $\tilde{\bf W}$. 

Let's work through a case of Metropolis dynamics, 
letting $W_{\alpha\bbeta} = w$ if $\Hclass(\alpha)< \Hclass(\beta)$, 
or $ w \exp[-\Hclass(\alpha)+\Hclass(\beta)]$ otherwise.  
Hence by \eqr{eq:Wsymm}, for $\beta\neq\alpha$, 
  \begin{equation}
      \tW_{\alpha\bbeta} = w \rme^{-\half |\Hclass(\alpha)-\Hclass(\beta)|}.
  \label{eq:WMetro}
  \end{equation}
The quantum hopping matrix element ($\thop$ term) must be proportional 
to $\tW_{\alpha\bbeta}$.
This depends only on the immediate environment of the flip location,
provided that $\Hclass$ is a sum of local terms. However, 
the $\thop$ term appears elaborate even in the simplest case, 
the model of Sec.~\ref{sec:spinmodel} on a one-dimensional chain
with  $\Hclass= - K \sum  S_i^z S_{i+1}^z$ . Flips that 
change $\Hclass$ are multiplied by $e^{-|K|/2}$.
Eq.~\eqr{eq:HFM} is replaced by
$\HHFM =   \sum _i - (\Jperp + \Jperp' S_{i-1}^z S_{i+2}^z)$
$(S_i^x S_{i+1}^x + S_i^y S_{i+1}^y)$
$- (\Jz+\Jz' S_{i-1}^z S_{i+2}^z)$ 
$ (S_i^z S_{i+1}^z - \frac{1}{4})$
where $\Jperp=t(1+e^{-|K|/2})$, 
$\Jz=V(1+e^{-|K|/2})$, 
$\Jperp'=t(1-e^{-|K|/2})$ and 
$\Jz'=V(1-e^{-|K|/2})$.

%%%  and the $\Vdim$ term is slightly worse.
\OMIT{E.g., in a lattice gas, the hopping term becomes modulated with 
the occupancy of neighboring sites.  
In the spin chain of Sec.~\ref{sec:spinmodel}, 
if the classical weight is according to
$\Hclass = \sum _i K S_i^z S_{i+1}^z$, then the exchange term in
\eqr{eq:HFM} becomes
$\exp(-\half K) (\cosh \half K + 4 \sinh \half K S_{i-1}^z S_{i+2}^z)
(S_i^+ S_{i+1}^- + S_i^- S_{i+1}^+) $...
The ``potential'' ($\Vdim$) term
disfavors the states with  higher classical energy
(all other things being equal!).}

\MEMO {{This is related to the above, but less important.
Consider it as an alternate text to be merged in somehow.}
We could make the same mapping at the Rokhsar-Kivelson point even 
for  a case in which the transition rate is different for different edges 
of the graph. (Notice that a flip of the same plaquette corresponds to many
different edges of the graph, depending on the configuration surrounding
it: thus, environment-dependent flip rates are included under this condition, 
provided each rate is symmetrically equal to the reverse flip rate.)
The potential cost is no longer proportional to the number of flip moves
in that state.
However, since we could vary the ratios between different flip rates, 
there is no longer a unique axis in parameter space to be varied as
we cross the RK point: in such models, the RK critical point sits in 
a multi-dimensional parameter space, and it is not obvious to me
whether the RK point is part of a line of critical points in most cases, 
or whether it tends to be a multicritical point in this enlarged space.}

\MEMO {The RK mapping
can never produce a quantum model with a sign problem, such as lattice fermions
or frustrated spin models.}

\section {Dynamic correlations}
\label{sec:dyn}

An amusing (perhaps useful) corollary of Eq.~\eqr{eq:classquantum}
is that for any generalized RK model, one can relate the quantum 
correlation function  
     \begin{equation}
         \Cquant_{BA}(\ttime) \equiv \la \hat{B}(\ttime) \hat{A}(0)\ra
     \end{equation}
to the similar classical one $\Cclass_{BA}(\ttime)$.
For the latter to make sense, the operators must be diagonal in Hilbert space, 
$\la \alpha |\hat{A}| \beta \ra = A_\alpha \delta_{\alpha\bbeta}$ and
similarly $\hat B$. 

Quite generally in a classical discrete system (starting in equilibrium)
     \begin{equation}
         \Cclass_{BA}(\ttime) = \sum _\alpha B_\bbeta p_\bbeta(\ttime|\alpha) A_\alpha
              \Pzero_\alpha ;
     \end{equation}
where $p_\bbeta(\ttime|\alpha)$ is the conditional probability, 
given that the state at $\ttime=0$ was $\alpha$ 
(which had probability $\Pzero_\alpha$).
First, $p_\bbeta(0|\alpha)=\delta_{\alpha\bbeta}$;
then \eqr{eq:timeevol} says $p_\bbeta(\ttime|\alpha)
=\sum c_\lambda \rme^{-\lambda \ttime} \philambda_\bbeta$; and orthonormality
of the eigenvectors (since $\bf W$ or $\tilde{\bf W}$
is symmetric) implies 
$c_\lambda= \philambda_\alpha$.  The latter formula
(with a {\it right} eigenvector) holds even for the case of 
Sec.~\ref{sec:weighted} where the master-equation matrix
$\bf W$ is nonsymmetric.
 
Using $\Pzero_\alpha = 1/\Ns$, the final result is
     \begin{equation}
         \Cclass_{BA}(\ttime) = 
         \sum _\lambda \rme^{-\lambda \ttime} \tA_\lambda \tB_\lambda 
     \label{eq:C-class}
     \end{equation}
where $\tA_\lambda \equiv {\Ns}^{-1/2} \sum _\alpha A_\alpha \philambda_\alpha$
(similarly $\tB_\lambda$). 

\OMIT{Notice that $\tA_0 \equiv \la A \ra _0$, so that as
$\ttime\to \infty$, eq.~\eqr{eq:C-class} indeed factors 
into $\la A \ra \la B \ra$.}

A quantum correlation function is computed using a similarity transform
by  the time-evolution operator to convert ${\hat B}(\ttime) \to
\rme^{\rmi\HH \ttime} {\hat B} \rme^{-\rmi\HH\ttime}$, an operator acting
at the same time as $\hat A$, and taking the expectation in the
ground state wavefunction:
\MEMO {NEED TO CHECK THE $\pm i$'s IN HERE.}
     \begin{equation}
         \Cquant_{BA}(\ttime) = 
       \la 0 |\rme^{\rmi \hat{\HH} \ttime} {\hat B} \rme^{-\rmi\hat{\HH}\ttime} 
       {\hat A} |0\ra . 
     \end{equation}
Insert a complete set of states $\sum |\lambda\ra \la \lambda|$ on
either side of the exponential between $\hat B$ and $\hat A$, 
and note that $\la \lambda | {\hat A} |0\ra = \tA_\lambda$
(similarly for $\hat B$). This equality follows from $|\lambda\ra =
\sum \tphilambda_\alpha |\alpha\ra$. (And it holds even in
the case that the state weights are unequal, Sec.~\ref{sec:weighted}.)
We obtain
     \begin{equation}
        \Cquant_{BA}(\ttime) = 
        \frac{1}{\Ns}
        \sum _\lambda \rme^{-\rmi (E_\lambda-E_0) \ttime} \tA_\lambda \tB_\lambda 
        = \Cclass_{BA}\Bigl(\frac{\rmi t}{\hbar w} \ttime \Bigr)
     \label{eq:Cclass-quant}
     \end{equation}
where I used \eqr{eq:E-lambda}.  Thus, the quantum correlations are
the {\it classical correlations in imaginary time}. An obvious application 
is that any dynamic  correlation,  measured via classical Monte Carlo 
with sufficient precision, may be converted
by analytic continuation to a quantum correlation function 
without the need to understand or compute the eigenfunctions. 
(Normally, for non-RK systems the same classical correlation function
must be obtained from a quantum Monte Carlo simulation, based on
a path integral and carried out in a space one dimension higher.)

It is fairly surprising to obtain the simple correspondence \eqr{eq:Cclass-quant}. 
It is true that the two evolution equations do have a 
corresponding time dependence, 
$\rme^{-\lambda \ttime}$ and $\rme^{-\rmi E_\lambda\ttime/\hbar}$
respectively, and the classical and quantum eigenfunctions are the same.
Yet the quantities which evolve according to these exponentials, 
and which are represented in the vector space spanned by those eigenfunctions, 
are {\it probability deviations} on the classical side, 
but {\it amplitudes} (which must be 
{\it squared} to obtain probabilities) on the quantum side.
\OMIT{The correspondence \eqr{eq:Cclass-quant} only
works because of the special fact that the probabilities (and amplitudes)
are the same for every microstate,  in the ground state (or the steady state). }

\ack
We acknowledge support by the National Science Foundation under
grant DMR-9981744.
\MEMO{CHECK NEW GRANT NO}
I thank Matthew Fisher for a comment about the dynamics.

%%%%%%%%%%%%%%%%%%%%%%%%%%%%%%%%%%%%%%%%%%%%%%%
%%
%% \begin{figure}[ht]
%%% \centering
%% \includegraphics[width=0.6\linewidth]{rkall.eps}
%% \caption{
%% TO COMPLETE. (a) dimer model (b) spin flips (c)
%% The state graph}
%% \label{fig:stategraph}
%% \end{figure}
%%
%%%%%%%%%%%%%%%%%%%%%%%%%%%%%%%%%%%%%%%%%%%%%%%


\section*{References}
\begin{thebibliography}{99}

\bibitem{qdm}  
D. Rokhsar and S. Kivelson,
Phys. Rev. Lett. 61, 2376 (1988).
%%% ``Superconductivity and the Quantum Hard-Core Dimer Gas"

\bibitem{Moe01a}
R. Moessner and S. L. Sondhi,
Phys. Rev. Lett 86, 1881 (2001).
%%% ``Resonating valence bond phase in the triangular lattice quantum dimer model''

\bibitem{Moe01b}
R. Moessner and S. L. Sondhi,
Phys. Rev. B 63, 224401 (2001).
%% ``Ising models of quantum frustration''.

\bibitem
{Na01} C. Nayak and K. Shtengel,
Phys. Rev. B 64, 064422 (2001),
%%% ``Microscopic models of two-dimensional magnets
%%% with fractionalized excitations''.

\bibitem
{Ba01} L. Balents, M. P. A. Fisher, and S. M. Girvin,  
%%% ``Fractionalization in an Easy-axis Kagome Antiferromagnet'',
Phys. Rev. B 65,  224412 (2002).  
%%% cond-mat/011005.

\bibitem{Mo02}
O. I. Motrunich and T. Senthil, 
%%% (cond-mat/0205170).
%%% ``Exotic order in simple models of bosonic systems''.
%%% Motrunich and  Senthil fractional; maybe others
Phys. Rev. Lett. 89, 27004 (2002).


\bibitem{fisher}
M. E. Fisher, Phys. Rev. 124, 1664 (1961);
%%% STATISTICAL MECHANICS OF DIMERS ON A PLANE LATTICE
P. Kasteleyn, Physica 27, 1209 (1961).
%%% STATISTICS OF DIMERS ON A LATTICE 
%%% 1. NUMBER OF DIMER ARRANGEMENTS ON A QUADRATIC LATTICE

\bibitem{fisher-st}
M. E. Fisher and J. Stephenson,
Phys. Rev. 132, 1411 (1963).
%% Correlations of dimers with monomers on a square lattice.

%% I decided to cut this -- I wrote the paper before seeing
%% any of theirs, and I don't belabor the quantum criticality 
%% \bibitem{ashvin}
%% ``Quantum Criticality and Deconfinement in Phase Transitions 
%% Between Valence Bond Solids
%% Authors: Ashvin Vishwanath, L. Balents, T. Senthil

\bibitem{dydim}
C. L. Henley, J. Stat. Phys. 89, 483 (1997).
%%% ``Relaxation time for
%%% a dimer covering with height representation'';
See Sec. I C.

\bibitem{statphys} C. L. Henley,
%%% ``Quantum and classical dynamics of the fully frustrated Ising model''.
abstract/talk at Statphys 20 (Paris, July 1998).

\bibitem{Bl02}
H. W. J. Bl\"{o}te and H. J. Hilhorst, J. Phys. A {15}, L631 (1982)
%% ``Roughening transitions and the zero-temperature
%% triangular Ising antiferromagnet''

\bibitem {Pa02} 
A. Paramekanti, L. Balents, and M. P. A. Fisher,
Phys. Rev. B 66, 054526 (2002).
%% ``Ring exchange, the Bose metal, and bosonization in two dimensions''.
%%% preprint (cond-mat/0203171),

\bibitem{q6v}
S. Chakravarty, Phys. Rev. B 66,  224505  (2002).
%% ``Theory of d-density wave viewed from a vertex model and its implications''
%.%% (cond-mat/0206282).
%%% ``Geometric frustration: Magic moments''
\bibitem{ramirez}
A. P. Ramirez, Nature 421, 483 (2003).


\bibitem{misguich} G. Misguich, D. Serban, and 
V. Pasquier, Phys. Rev. Lett. 89, 137202 (2002). 
%%% Quantum dimer model on the Kagome lattice:
%% solvable dimer-liquid and Ising gauge theory
%%% omitted G. Misguich, D. Serban, and V. Pasquier, 
%%% Phys. Rev. B 67, 214413 (2003) 
%%% quantum dimer model with extensive ground state
%%% entropy on the Kagome lattice. 

\bibitem{hermele} M. Hermele and M.~P.~A. Fisher, 
preprint (cond-mat/0305401). 

\bibitem{zeng}
C. Zeng and C. L. Henley
Phys. Rev. B 55, 14935-47 (1997).
%%% ``Zero-temperature phase transitions of an antiferromagnetic Ising
%%% model of general spin on a triangular lattice''

\bibitem{qdspin}
C. L. Henley, 
``Spin dynamics from the quantum dimer model'',
to be submitted.
%% (QD-spin). 

\bibitem {fazekas}
P. Fazekas and P. W. Anderson,
Phil. Mag 30, 423-440 (1974);
B. Kleine, P. Fazekas, and E. M\"uller-Hartmann,
Z. Phys. B 86, 405-10 (1992).

\bibitem{vB77} 
H. van Beijeren, Phys. Rev. Lett. 38, 993 (1977).
%% ``Exactly Solvable Model for the Roughening Transition of a Crystal Surface''.


%%% \bibitem
%%% {An56} P. W. Anderson,
%%% Phys. Rev. 102, 1008-13 (1956),
%%% ``Ordering and antiferromagnetism in ferrites''.
%%% pyrochlore

%%% \bibitem {Li86} R. Liebmann,
%%% {\it Statistical Physics of Periodically Frustrated Ising Systems}
%%% (Springer Lecture Notes on Physics No. 251, 1986).





\end{thebibliography}
\end{document}